\documentclass[preprint,11pt]{elsarticle}

\usepackage{amssymb}
\usepackage{amsmath}
\usepackage{natbib}
\usepackage{hyperref}
\usepackage{graphicx}
\graphicspath{{Figures/}}
\usepackage{xcolor}
\usepackage{lineno}
\modulolinenumbers[5]

\begin{document}
\begin{frontmatter}

\title{Presurgical Neural Energy Landscapes Predict Postoperative Working Memory Outcome After Brain Tumor Resection}

\author[1]{Triet M. Tran}
\author[1,2]{Sina Khanmohammadi\corref{cor}}

\affiliation[1]{organization={University of Oklahoma},
            addressline={Data Science And Analytics Institute},
            city={Norman},
            postcode={73109},
            state={Oklahoma},
            country={USA}}
            
\affiliation[2]{organization={University of Oklahoma},
            addressline={School of Computer Science},
            city={Norman},
            postcode={73109},
            state={Oklahoma},
            country={USA}}

\cortext[cor]{Corresponding Author (sinakhan@ou.edu)}

\begin{abstract}
Surgical resection is the primary treatment option for brain tumor patients, but it carries the risk of postoperative cognitive impairments. This study investigates how tumor-induced alterations in presurgical neural dynamics relate to postoperative working	memory outcome assessed by Spatial Span (SSP) test. We analyzed functional magnetic resonance imaging (fMRI) of brain tumor patients before surgery and extracted energy landscapes of high-order brain interactions. We then examined the relation between these energy features and postoperative working memory performance using statistical and machine learning (random forest) models. Patients with lower postoperative SSP Scores (2 to 5) exhibited fewer but more extreme transitions between local energy minima and maxima, whereas patients with higher SSP Scores (6 to 9) showed more frequent but less extreme shifts. Furthermore, the presurgical high-order energy features were able 	to accurately predict postoperative working memory outcome with a mean accuracy of 90\%, F1 score of 87.5\%, and an AUC of 0.95. Our study suggests that the brain tumor-induced disruptions in high-order neural dynamics before surgery are predictive of postoperative working memory outcome. Our findings pave the path for personalized surgical planning and targeted interventions to mitigate cognitive risks associated with brain tumor resection.
\end{abstract}


\end{frontmatter}

\section{Introduction}
\label{sec:introduction}
Brain tumor resection is the primary treatment option for brain cancer \cite{owonikoko2014current}. The process involves removing as much of the tumor as possible while minimizing damage to surrounding healthy brain tissue \cite{rodriguez2024innovations}. This procedure is crucial for alleviating symptoms, improving neurological function, and extending the patient's life \cite{young2023surgical}. However, it is also often associated with postoperative cognitive impairments \cite{aldape2019challenges}, which could persist long-term and significantly impair patients’ quality of life \cite{taphoorn2004cognitive}. Hence, Identifying presurgical markers associated with postoperative cognitive outcomes could enhance patient management and personalize postoperative care \cite{kirkman2023systematic}.

Several studies have highlighted potential predictors of postoperative cognitive outcomes in brain tumor patients, ranging from the tumor's location, size, and type, to the patient's age, pre-existing cognitive function, and overall health \cite{dallabona2017impact,rijnen2019cognitive,schiavolin2021outcome,zangrossi2022presurgical,tariq2025cognitive}. Nonetheless, each patient's condition is unique, and the interplay between these predictors can vary significantly, making it difficult to pinpoint which factors are most influential in any given case \cite{dadario2021reducing}. Recent studies have explored the applicability of neuroimaging in predicting cognitive outcome after brain tumor resection \cite{lakhani2023current,gu2024seeing,griffiths2025limited}. One of the key factors that seems to be emerging from these studies is the role of functional connectivity in post-surgical cognitive decline \cite{wang2022characterization,herbet2024predictors}. For instance, the functional connectivity of resected neural tissue has been found to correlate with neurological morbidity \cite{tarapore2012magnetoencephalographic}, reduced connectivity in the parietal region of the non-tumor hemisphere is linked to poor neuropsychological outcomes \cite{lang2017functional}, variability in information processing pressure on functional hub regions has been shown to correlate with post-operation verbal memory \cite{carbo2017dynamic}, and the functional connectivity within the Default Mode Network (DMN) and the bilateral frontoparietal regions has been shown to correlate with cognitive performance in meningioma patients \cite{van2018cognitive}. More recently, Luckett and colleagues have shown that functional connectivity and the tumor's location relative to specific networks are strong predictors of brain surgery outcomes, whereas tumor volume is only a moderate predictor \cite{luckett2024predicting}.

The majority of previous studies have focused on the direct analysis of low-order functional connectivity measures and their correlation with post-surgery cognitive outcomes. Low-order functional connectivity measures mainly focus on the correlation of signals from different brain regions \cite{bastos2016tutorial,fox2018functional}. However, several studies suggest that high-order functional interactions between groups of regions could provide valuable information that is not accessible through low-order connectivity measures \cite{herzog2022genuine,santoro2024higher}. In our recent work, we have shown that taking into account higher-order functional connectivity measures could significantly increase the predictive power of presurgical fMRI recordings for estimating post-surgery working memory outcome \cite{tran2024high}. Furthermore, most of the existing studies have considered static functional connectivity measures, which give an overall snapshot of brain interactions. However, many studies highlight the importance of dynamics in functional connectivity and its relation to cognitive function \cite{braun2015dynamic,preti2017dynamic}, for example, recent work characterizes dFC within a low-dimensional 
state-space and shows that transitions in a compact state space provide a general framework linking neural flexibility and neuroplasticity to cognitive status \cite{razavi2025brain}.

Given these findings, we explored the presurgical high-order brain dynamics extracted from fMRI data and their predictive power for post-surgery working memory outcome. Specifically, we conducted a high-order energy landscape analysis on a sample-by-sample basis and examined its correlation with post-surgery working memory performance. The energy values defined for each brain state represent the stability of that state, with lower energy values indicating more stable states. Therefore, our analysis aimed to capture the interdependency of neural stability with post-surgery working memory outcomes. Our results show a significant correlation between presurgical energy landscape fluctuations and post-surgery working memory performance in brain tumor patients. Furthermore, we analyzed the predictive power of these energy landscape features within a machine learning framework, showing that the energy landscape features could accurately predict post-surgery working memory outcomes. Together, these results suggest a strong link between pre-surgery high-order neural dynamics and post-surgery cognitive outcomes in brain tumor patients. Our results could help healthcare providers tailor rehabilitation programs and implement strategies to mitigate cognitive deficits, improving overall patient care and recovery.
\section{Methods}

\subsection{Data Description}
\subsubsection{Study Participants}
The data for this study were obtained from a publicly available dataset\cite{aerts2022pre}. In brief, the original data collection took place at Ghent University Hospital (Belgium) between May 2015 and October 2017, where adult patients ($\geq 18$ years of age) with either a supratentorial glioma (WHO grade II or III) or meningioma (WHO grade I or II) were recruited. All participants provided informed consent in accordance with the Declaration of Helsinki and local Ethics Committee guidelines. The original contains 25 tumor patients and 11 healthy controls. In the present study, we considered a subset of 20 tumor patients (7 with gliomas and 13 with meningiomas) who underwent resting-state functional MRI (rs-fMRI) scans and cognitive testing both before and six months after surgical resection. Subjects without complete rs-fMRI scans or cognitive assessments were excluded from the analysis. Patients met inclusion criteria if they were capable of completing neuropsychological evaluations and were medically approved for MRI examination. 

\subsubsection{Image Acquisition}
Neuroimaging data were acquired on a 3T Siemens Magnetom Trio scanner with a 32-channel head coil, following the data acquisition protocol outlined in \cite{aerts2022pre}. T1-weighted (T1w) anatomical images were collected using a magnetization-prepared rapid gradient-echo (MPRAGE) sequence (TR $= 1750$~ms, TE $= 4.18$~ms, voxel size $1 \times 1 \times 1 \,\text{mm}$), ensuring high-resolution structural coverage. Resting-state fMRI scans were obtained using an echo-planar imaging sequence (TR $= 2100$--2400~ms, TE~$= 27$~ms, voxel size $3 \times 3 \times 3 \,\text{mm}$), lasting approximately 6--7~minutes in total. During rs-fMRI, participants were instructed to keep their eyes closed, remain still, and stay awake. 

\subsubsection{Data Preprocessing}
The rs-fMRI data were available as preprocessed time series data, following the pipeline described in \cite{aerts2022pre}. We did not apply any additional preprocessing beyond what was originally performed. The original preprocessing steps briefly included: discarding the first few fMRI volumes to allow magnetization equilibrium, followed by head motion correction and slice-timing correction. Non-brain tissues were removed with BET, and spatial normalization was performed by registering each participant's functional images to their T1w structural image and then warped to MNI standard space. High-pass filtering using a 100-second cutoff (equivalent to $0.01$ Hz) was applied to reduce low-frequency drifts. Tumor regions were segmented or masked in the structural data ensuring that large mass effects or infiltrations did not compromise registration or parcellation. All registrations were visually inspected, focusing on peri-tumoral regions for potential misalignment. 
	
\subsubsection{Working Memory Assessment}
Working memory was measured using the Spatial Span (SSP) test, with raw scores ranging from 2 to 9. Following our previous study\cite{tran2024high}, these scores were grouped into ``low” (2-5) versus ``high” (6-9) working memory performance. Patients were assessed both before and six months after tumor resection. In this study, we focused on postoperative outcomes, as they reflect the functional impact of surgical intervention and are most relevant to clinical recovery. We specifically focused on working memory assessment due to its well-documented clinical persistence in brain tumor patients \cite{amores2025neuropsychological}, its relevance to probing the neural energy landscape framework \cite{finc2020dynamic}, and requiring minimal verbal or complex motor responses, making it suitable for assessing post operation clinical population \cite{diaz2022challenge}.

\subsection{High-order Energy Landscape Framework}
\begin{figure}[htbp]
\centering
\includegraphics[width=\linewidth]{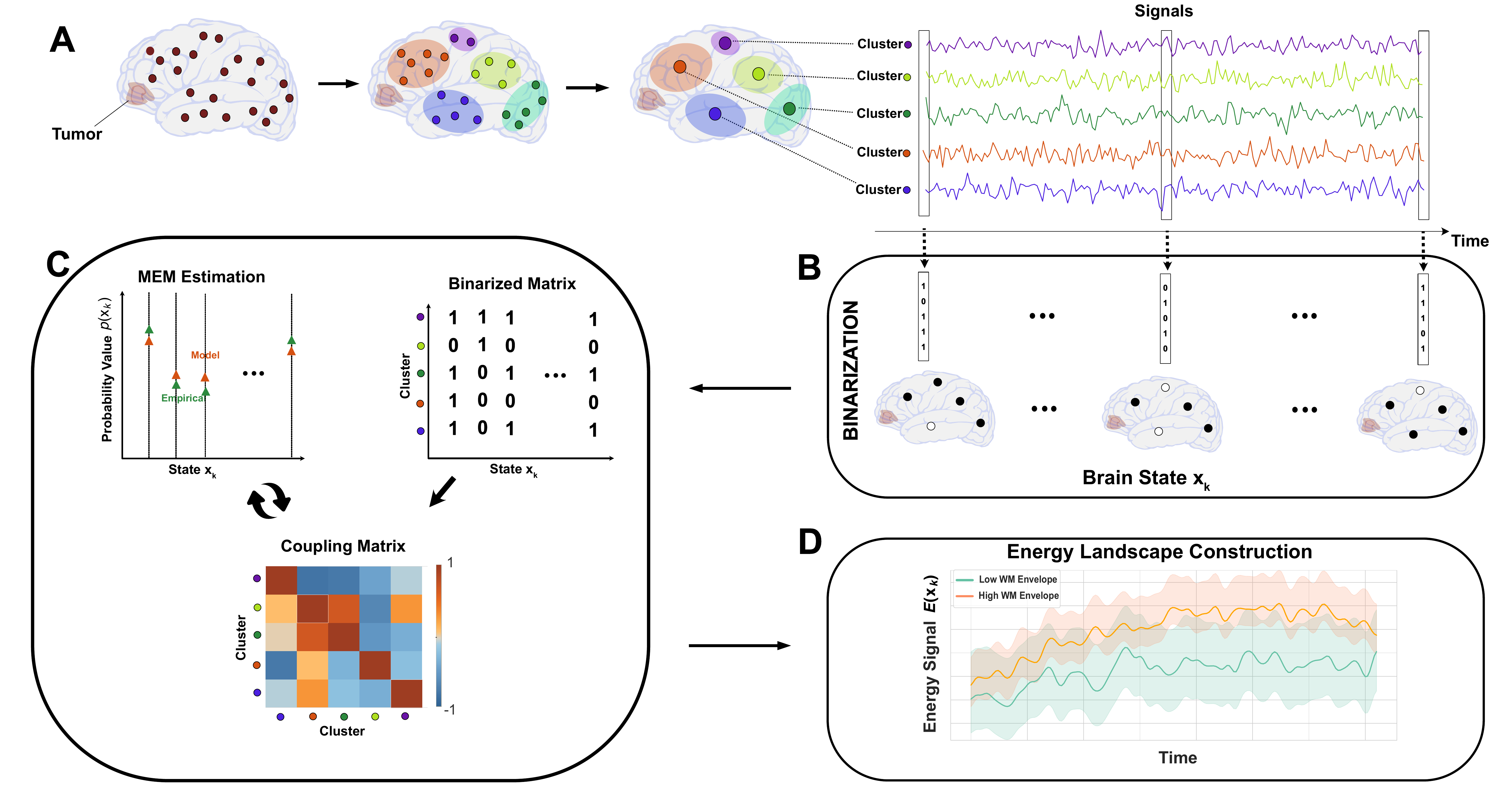}
\caption{\textbf{High-order energy landscape framework.} 
\textbf{(A)}~\textit{Identifying Functional Clusters:} Resting-state fMRI signals are grouped into functionally similar clusters using \textit{k}-means algorithm. Each cluster represents a group of brain regions with similar activity. 
\textbf{(B)}~\textit{Extracting Brain States:} The average signal from each cluster is converted into binary states (active = 1 and inactive = 0) based on a threshold (e.g. mean value), forming a sequence of brain states over time. 
\textbf{(C)}~\textit{Fitting Maximum Entropy Model:} The maximum entropy model is fitted to identify the probability distribution of brain states that maximizes entropy, while ensuring that the model matches the observed average activation of each cluster and the pairwise co-activation between clusters.
\textbf{(D)}~\textit{Constructing Energy Landscapes:} 
The parameters of the resulting optimum probability distribution is utilized to calculate the energy value of each state.  
}
\label{fig:tp}
\end{figure}

The energy landscape method presented in \cite{kang2019graph} enables the analysis of energy fluctuations between brain states on a sample-by-sample basis, offering insights into the dynamics of brain state transitions. Nonetheless, the original method considers a low-order description of brain states, where each state is defined based on the fMRI signals from individual regions of interest (ROIs). Here, we propose an extension of this approach, where each state is described using a cluster of ROIs to account for the high-order interactions between different brain regions. The overall framework (as shown in Figure~\ref{fig:tp}) consists of four main steps, which are described in the following sections. The notations used throughout these sections are summarized in Table~\ref{tab:notation}.

\begin{table}[htbp]
\caption{Summary of notation used throughout the paper}
\begin{tabular}{|p{50pt}|p{290pt}|}
\hline
Symbol & Description
\\
\hline
$N$ & Total Number of Brain Region Clusters. \\
$t$ & The Time Index. \\
$\mathbf{x}_k$ & Binary State Vector Corresponding to the $k$-th Brain Configuration.\\
$x_{ki}$ & Activation State (0 or 1) of Cluster $i$ in $k$-th Brain Configuration.\\
$\langle x_i\rangle_{\text{data}}$ & Empirical Mean Activation of Cluster $i$.\\
$\langle x_i x_j\rangle_{\text{data}}$ & Empirical Pairwise Correlation between Clusters $i$ and $j$.\\
$\langle x_i\rangle_{\text{model}}$ & Derived Mean Activation of Cluster $i$.\\
$\langle x_i x_j\rangle_{\text{model}}$ & Derived Pairwise Correlation between Clusters $i$ and $j$.\\
$p(\mathbf{x}_k)$ & Probability of the Brain State Configuration $\mathbf{x}_k$.\\
$h_i$ & Intrinsic Activation Bias (Baseline Activation Tendency) for Cluster $i$.\\
$\mathbf{W}_{ij}$ & Coupling Strength between Clusters $i$ and $j$.\\
$E(\mathbf{x}_k)$ & Energy Associated with Brain Configuration $\mathbf{x}_k$.\\
$S$ & Total Number of Selected Features in the Random Forest Model.\\
\hline
\end{tabular}
\label{tab:notation}
\end{table}

\subsubsection{Identifying Functional Clusters}
The first step of the proposed framework involves identifying the functional clusters where brain regions in each cluster have similar activation patterns (Figure~\ref{fig:tp}.A). In order to achieve this, we performed $k$-means clustering on the entire set of cortical and subcortical regions for each subject, grouping ROIs based on the similarity of their BOLD signal patterns. The total number of brain  clusters $N$ was selected using the elbow method, which identifies the point where increasing $N$ yields diminishing improvement in within-cluster variance, ensuring a balance between model simplicity and functional resolution. Each cluster's mean time series was extracted for subsequent analyses. This data-driven approach allowed us to identify high-order functional interactions that are not constrained by predefined network structures.

\subsubsection{Extracting Brain States}
The extracted continuous signals from each cluster $i$ were set to a discrete state of active ($x_i=1$) or inactive ($x_i=0$) at each time point $t$. We employed a simple thresholding method where the values above the mean were considered active and values below were set as inactive. This approach is conceptually similar to median-based binarization that has been used in previous studies\cite{watanabe2013pairwise}. We selected a mean-based threshold for simplicity and consistency. Empirically, this yields activation ratios in the range of 30–70\%, ensuring that the signals are neither too sparse nor too saturated\cite{watanabe2014energy}.

\subsubsection{Fitting Maximum Entropy Model}
Once the time series data were binarized, a maximum entropy model (MEM) is fitted to estimate the probability distribution of the brain states that maximizes entropy while preserving the observed first-order (mean activation) and second-order (pairwise correlations) statistics \cite{kang2017energy}. The maximum entropy model is based on the principle that, given the constraints of the observed data, the probability distribution that best represents the current state of a system is the one with the largest entropy (randomness) \cite{sutter2019generalized}. The MEM estimates the probability distribution $p(\mathbf{x}_k)$ over all $2^N$ possible states by solving the following constrained optimization problem:

\begin{equation}
\begin{aligned}
    &\text{maximize:}\quad -\sum_{k=1}^{2^N} p(\mathbf{x}_k)\,\log p(\mathbf{x}_k) \\
    &\text{subject to:}\quad \langle x_i \rangle_{\text{model}} =  \langle x_i \rangle_{\text{data}} \\
    & \quad \quad \quad \quad \quad \langle x_i x_j \rangle_{\text{model}} = \langle x_i x_j \rangle_{\text{data}},
\end{aligned}
\label{eq:mem_opt}
\end{equation}

where $\mathbf{x}_k=(x_{k1},\ldots,x_{kN})$ represents the binary state vector corresponding to the $k$-th brain configuration. $\langle x_i \rangle_{\text{data}}$ represents the empirically observed mean activations, $\langle x_i x_j \rangle_{\text{data}}$ shows the empirically observed pairwise correlation, $\langle x_i\rangle_{\text{model}}$ corresponds to derived mean activation, and $\langle x_i x_j\rangle_{\text{model}}$ stands for the derived pairwise correlation between clusters $i$ and $j$. 

This optimization problem could be solved by applying Lagrange multipliers $h_i$ and $\mathbf{W}_{ij}$ to incorporate the linear constraints imposed by the empirical statistics, where the optimal parameters are estimated using gradient ascent method. The resulting probability values form a Boltzmann distribution in the following form \cite{jaynes1957information}:

\begin{equation}
p(\mathbf{x}_k) = \frac{\exp[-E(\mathbf{x}_k)]}{\sum_{k=1}^{2^N}\exp[-E(\mathbf{x}_k)]},
\end{equation}

where the energy function $E(\mathbf{x}_k)$ is given by:

\begin{equation}
E(\mathbf{x}_k) = -\sum_{i<j} \mathbf{W}_{ij}\, x_{ki} x_{kj} - \sum_{i} h_i\, x_{ki}.
\label{eq:energy}
\end{equation}

The parameters $h_i$ and $\mathbf{W}_{ij}$ here correspond to the baseline activation bias of cluster $i$ and coupling strength between clusters $i$ and $j$, respectively. As mentioned previously, these parameters act as Lagrange multipliers to incorporate the constraints in equation \ref{eq:mem_opt}, ensuring that the resulting distribution matches the observed first and second-order statistics of the data. We considered a fit to be acceptable when the accuracy of fit, defined as the correlation between model-predicted and empirical statistics, exceeded $0.8$, as suggested by Watanabe\cite{watanabe2014energy}. Hence, we only considered the coupling matrices ($\mathbf{W}_{ij}$) from models meeting this criterion for further analysis.

\subsubsection{Constructing Energy Landscapes}
After identifying the optimal parameters $\{h_i, \mathbf{W}_{ij}\}$, we computed the energy $E(\mathbf{x}_k)$ for each possible configuration $\mathbf{x}_k$ using Eq.\eqref{eq:energy}. Lower energy states occur more frequently and represent stable brain configurations, whereas higher energy states are less probable and can be viewed as transitional states that require more ``energy'' to maintain\cite{kang2019graph}. The full distribution of energy values across all configurations constitutes the energy landscape, highlighting valleys of stability (low energy) and peaks of transition (high energy). By examining these energy landscapes, we gain insights into how the brain navigates between more stable and less stable functional states over time, potentially elucidating differences associated with different cognitive performances.

\subsection{Statistical Analyses and Predictive Modeling}
After obtaining the energy landscape values, we conducted several statistical and machine learning analyses to evaluate the relationship between high-order energy values and the working memory outcome in brain tumor patients. 

\subsubsection{Group Differences Analysis}
We first examined high-order versus low-order energy landscape features and their relationship to postoperative working memory performance. The low-order energy landscape was computed in the same manner as the high-order model; however, instead of using clusters of regions of interest (ROIs), we selected individual ROIs from canonical resting-state networks (RSNs). These included four well-characterized networks: the Default Mode Network (DMN), Salience Network (SN), Sensorimotor Network (SMN), and Limbic Network (LN). After extracting the energy landscapes for both the high-order and low-order models, we identified all local minima and maxima by comparing each state's energy to that of its neighboring states. We then quantified the following:
\begin{itemize}
\item The \emph{energy envelope}, defined as the smoothed time series connecting local minima and maxima across the energy landscape. 

\item The distribution of extreme energy values, defined as the highest and lowest 20\% of energy values within each subject, where the subject-level summaries were then compared between groups. We chose the 20\% threshold to focus on the most energetically distinct states while avoiding dilution by more moderate values. Similar thresholds have been used in prior neuroimaging studies to emphasize extreme states in neural dynamics \cite{watanabe2014energy, ezaki2017energy}.

\item The \emph{frequency} of transitions between minima and maxima energy values, reflecting the brain's flexibility in transitioning between states. 
\item The \emph{magnitude} of transitions, capturing the energy difference between local minima and maxima, indicating the relative cost of transitioning between states. 
\end{itemize}

We then conducted group-level statistical comparisons using the Mann-Whitney \textit{U} test to examine whether metrics derived from the energy landscapes (e.g., distribution of local minima and maxima, frequency of transitions, and magnitude of transitions) differ between the high vs low performing cohorts. All tests were two-tailed with significance set at $p < 0.05$.

\subsubsection{Random Forest Analysis}
To evaluate the predictive power of the extracted energy landscape features, we trained a random forest classifier to identify the high vs low postoperative working memory performance. Separate models were trained based on the high-order energy features (extracted from clusters of ROIs), low-order energy features (extracted from canonical networks), and conventional functional connectivity (FC) values (calculated as Pearson correlations between normalized time series signals).

Model performance was evaluated using leave-one-out cross-validation (LOOCV). Because random forest training and hyperparameter selection are sensitive to random seed initialization, we repeated the analysis 30 times with different random seeds and summarized performance metrics across repetitions. This number of replications was chosen to balance the computational cost and reliability of performance estimates. In practice, the choice of number of replications could influences the precision of p-value estimates and therefore the p-values in ur results should be considered approximates interpreted accordingly.

We also trained a separate set of models using the combination of low-order energy features from each canonical network with high-order energy features to assess the relative importance of each feature set. For each model, we computed feature importance scores using the mean decrease in Gini impurity. We focused on the top $S$ most important features, where $S$ ranged from 2 to 10. For each $S$, we calculated the proportion of high-order versus low-order features and compared the corresponding contributions. 

\section{Results}

\subsection{High-Order Energy Landscape Features Reveal Significant Differences in Patients with High vs Low Working Memory Performance}
\label{sec:result:first}

\begin{figure}[htbp]
\centering
\includegraphics[width=0.9\textwidth]{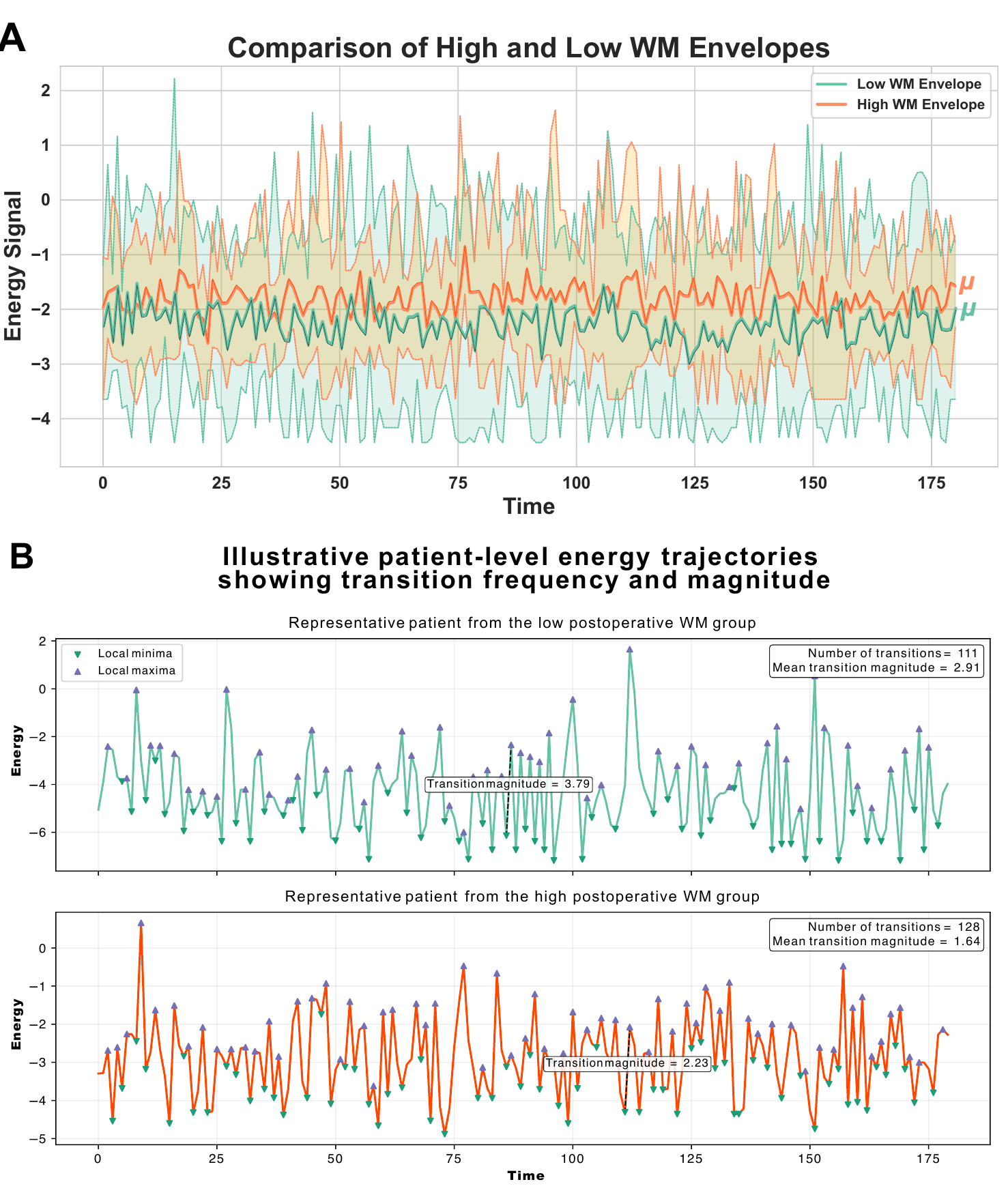}

\caption{%
\textbf{Energy envelope comparison between low and high working memory (WM) groups.}
(\textbf{A}) Group-level energy envelopes showing the amplitude range of energy fluctuations across time for patients in the low WM (green) and high WM (orange) groups. Patients with low WM exhibit larger-amplitude fluctuations and broader excursions into lower-energy states, whereas the high WM group demonstrates more constrained energy fluctuations. The solid lines represent the mean energy signals for each group. 
(\textbf{B}) Representative patient-level energy trajectories from the low and high postoperative WM groups, illustrating how local minima and maxima were identified and how transition frequency and transition magnitude were quantified over time.
}
\label{fig:envelope}
\end{figure}

We first visualized the overall energy signal patterns by plotting the energy envelopes of patients in the high and low working memory (WM) groups. As shown in Figure~\ref{fig:envelope}.A, the energy envelopes of the two groups exhibit different temporal dynamics and amplitude ranges. The low WM group shows larger-amplitude fluctuations and broader excursions into low-energy states, whereas the high WM group exhibits more constrained energy fluctuations across time. Representative patient-level trajectories consistent with these group-level patterns are shown in Figure~\ref{fig:envelope}.B, illustrating how transition frequency and transition magnitude were defined over time.

\begin{figure}[htbp]
\centering
\includegraphics[width=\textwidth]{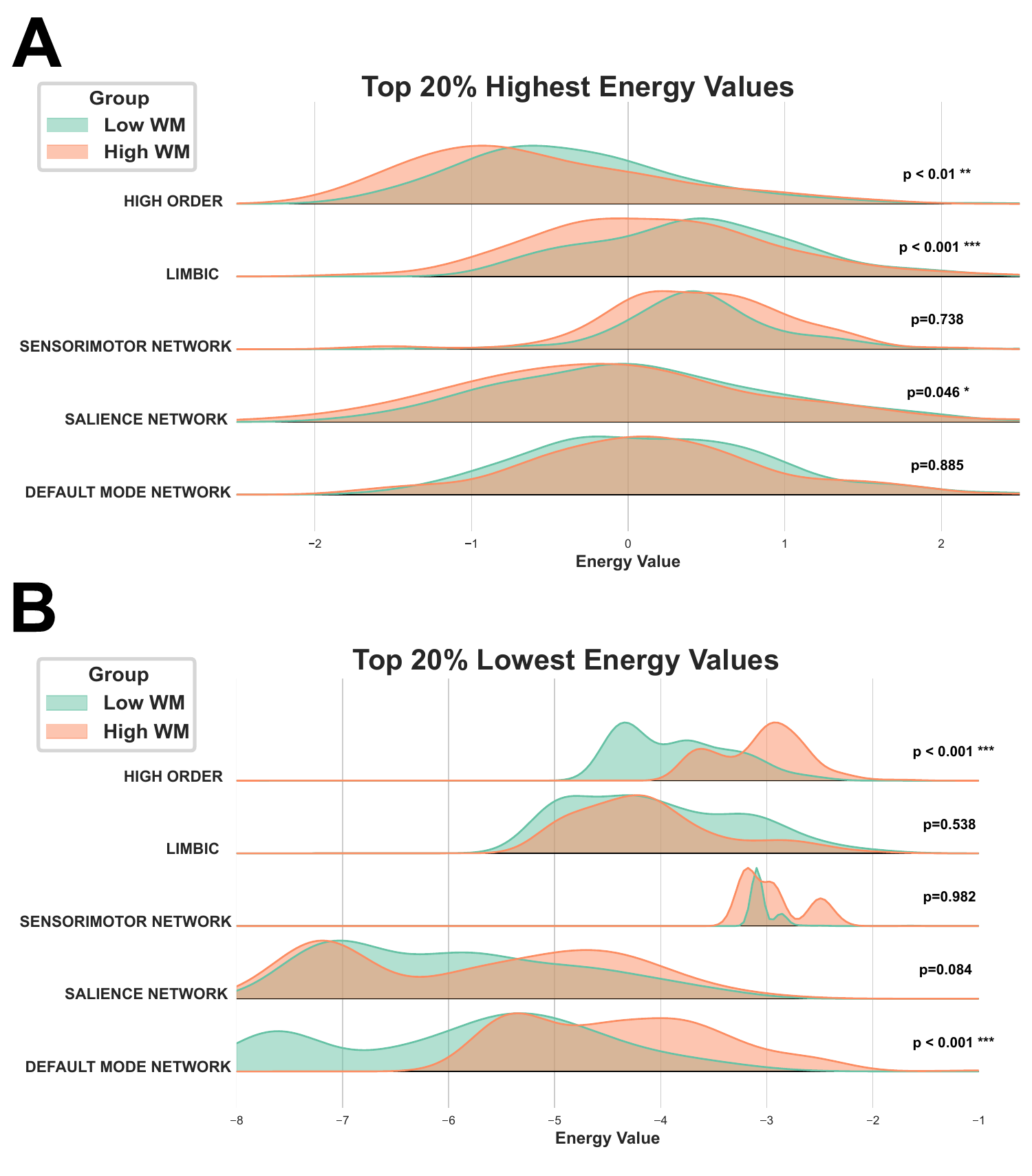}
\caption{%
\textbf{Distributions of energy values across canonical and high-order networks.}
Distributions of the (\textbf{A}) top $20 \%$ \emph{highest} and (\textbf{B}) top $20 \%$ \emph{lowest} energy values are shown for the default mode, salience, sensorimotor, limbic, and high-order networks. 
Values are compared between low (green) and high (orange) working memory groups, with $p$-values from Mann-Whitney \textit{U} tests displayed for each comparison.
}
\label{fig:energy_metrics}
\end{figure}

In order to quantify these differences, we compared extreme energy values extracted from the high-order network and the four canonical resting-state networks (default mode, salience, sensorimotor, and limbic). As shown in Figure~\ref{fig:energy_metrics}.A, the top $20 \%$ highest energy values were significantly different in the high-order network ($p=0.002$) and two of the canonical networks (salience: $p=0.046$ and limbic: $p < 0.001$), whereas no significant group differences were observed in the default mode ($p=0.885$) or sensorimotor ($p=0.738$) networks. The examination of the top $20 \%$ lowest energy values (Figure~\ref{fig:energy_metrics}.B) revealed robust group differences for both the high-order ($p < 0.001$) and the default mode networks ($p < 0.001$). No statistically significant effects were observed in the salience ($p=0.084$), sensorimotor ($p=0.982$) or limbic ($p=0.538$) networks. Taken together, these results indicate that high-order energy features consistently capture the differences in neural dynamics between low and high working memory performance.

\begin{figure}[htbp]
\centering
\includegraphics[width=\textwidth]{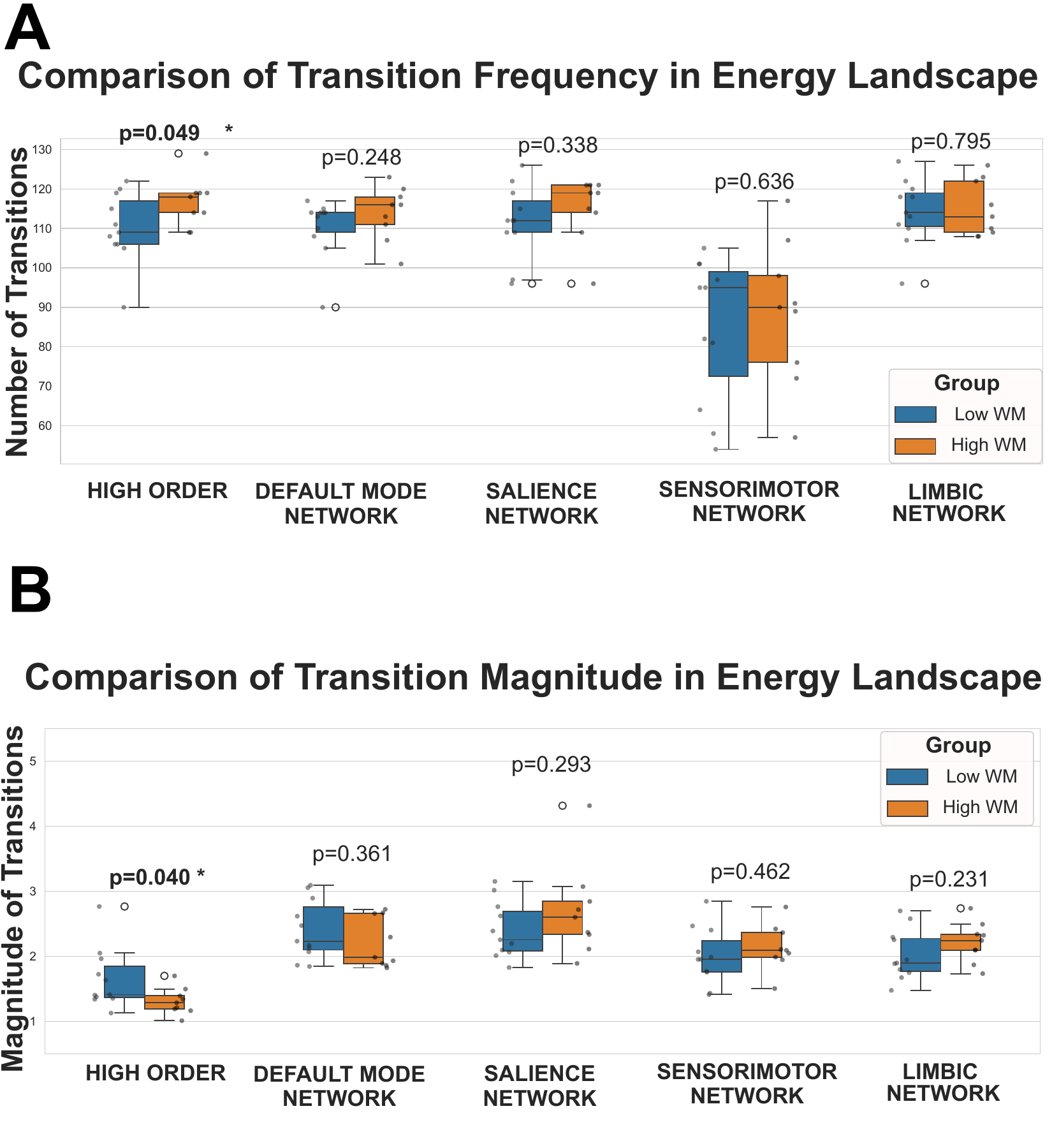}
\caption{%
\textbf{Comparison of energy transition features for low vs.\ high working memory (WM) groups.}
(\textbf{A})~Boxplots showing the number of transitions between local minima and maxima in the energy landscapes derived from the high-order approach (\textit{k}-means) and four canonical networks (default mode, salience, sensorimotor, limbic). 
(\textbf{B})~Boxplots of the transition magnitude, defined as the absolute difference in energy when moving from a local minimum to its adjacent local maximum. 
Mann-Whitney \textit{U} test \textit{p}-values are provided above each pair of boxplots.
}
\label{fig:second_result}
\end{figure}

Next, we examine whether the frequency and magnitude of transitions in the energy landscapes could differentiate patients with low versus high working memory performance. As shown in Figure~\ref{fig:second_result}.A, high-order energy landscape features revealed a significant difference in the number of transitions between the two groups ($p = 0.049$), whereas all four canonical networks (default mode, salience, sensorimotor, and limbic) displayed non-significant differences. A similar pattern emerged when assessing the magnitude of transitions (Figure~\ref{fig:second_result}.B). The high-order analysis showed a statistically significant group difference ($p=0.040$), while the canonical network approaches did not show any significant differences ($p>0.05$) between the low and high working memory groups. These findings suggest that high-order energy landscape analysis may provide valuable insights into the underlying neural dynamics that support working memory in patients with brain tumors.

\subsection{High-Order Energy Landscape Features are Predictive of Post-Surgery Working Memory Performance}
\label{sec:results:third}

\begin{figure}[htbp]
\centering
\includegraphics[width=\textwidth, height=15.4cm, keepaspectratio]{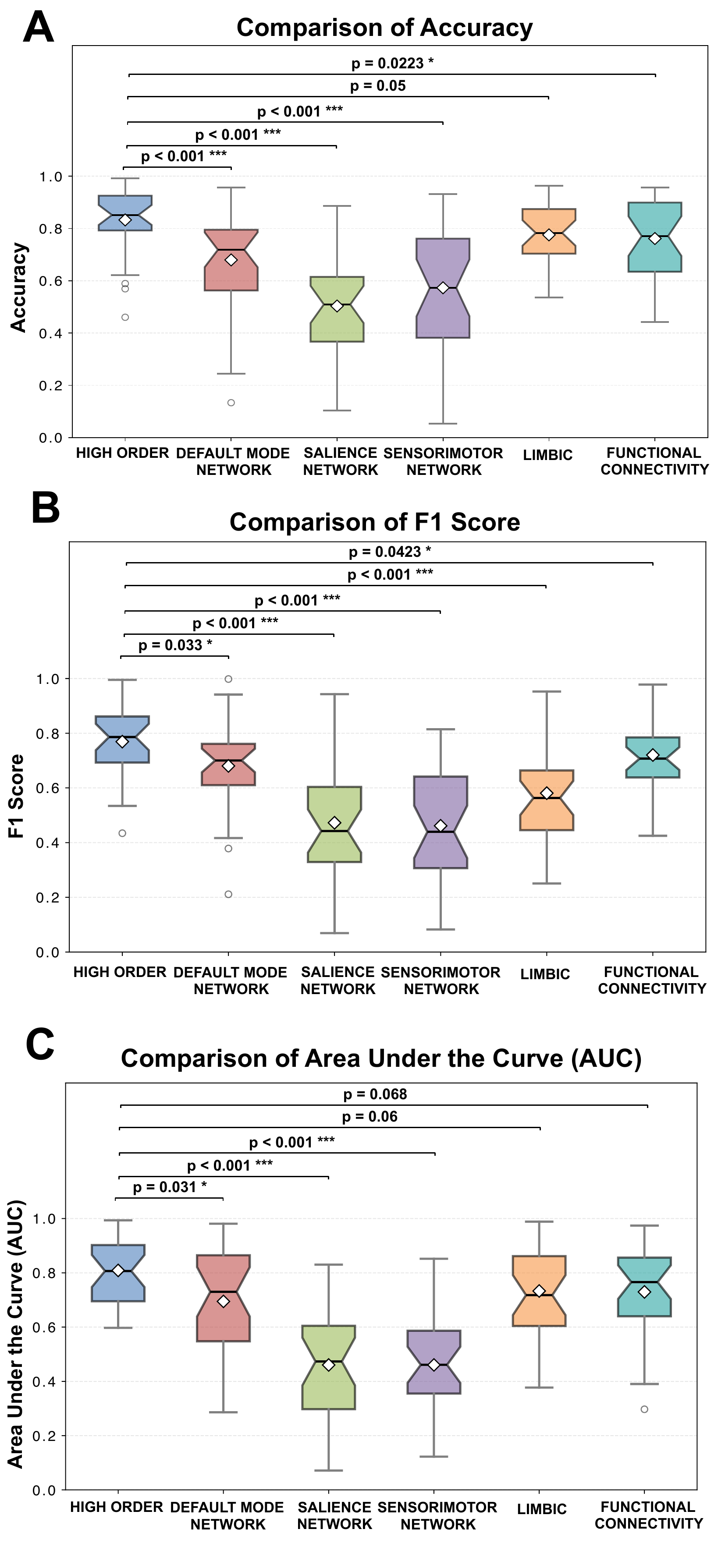} 
\caption{%
\textbf{Classification performance based on random forest models.}
The figure presents the distribution of (\textbf{A}) Accuracy, (\textbf{B}) F1 Score, and (\textbf{C}) Area Under the Curve (AUC) obtained from random forest classifiers trained with six different feature sets: high-order, default mode, salience, sensorimotor, limbic, and functional connectivity. Each boxplot summarizes results from 30 repeated runs under the same leave-one-out cross-validation framework. Statistical significance between the high-order feature set and each of the other feature sets was assessed using pairwise Mann-Whitney U-tests, with $p$-values annotated above each comparison.}
\label{fig:acc_f1_auc}
\end{figure}

The performance of random forest models trained on high-order energy landscape features is presented in Figure~\ref{fig:acc_f1_auc}. The high-order model achieved a mean accuracy of $90\%$, an F1 score of $87.5\%$, and an AUC of $0.95$. By contrast, models based on low-order networks exhibited significantly lower mean accuracy (below $80\%$), lower F1 scores (ranging from $47.1-70.6\%$), and lower AUC values ($0.47$ to $0.78$). Statistical comparisons (Mann-Whitney U-tests) confirmed that the high-order model significantly outperformed the low-order models across all evaluated metrics (accuracy, F1 score, and AUC), except for the limbic network. We also included a conventional functional connectivity (FC)-based model as a baseline, which demonstrated significantly lower predictive performance (mean accuracy below $80\%$) compared to the high-order energy model, highlighting the added value of energy landscape features.

\begin{figure}[htbp]
    \centering
    \includegraphics[width=\textwidth]{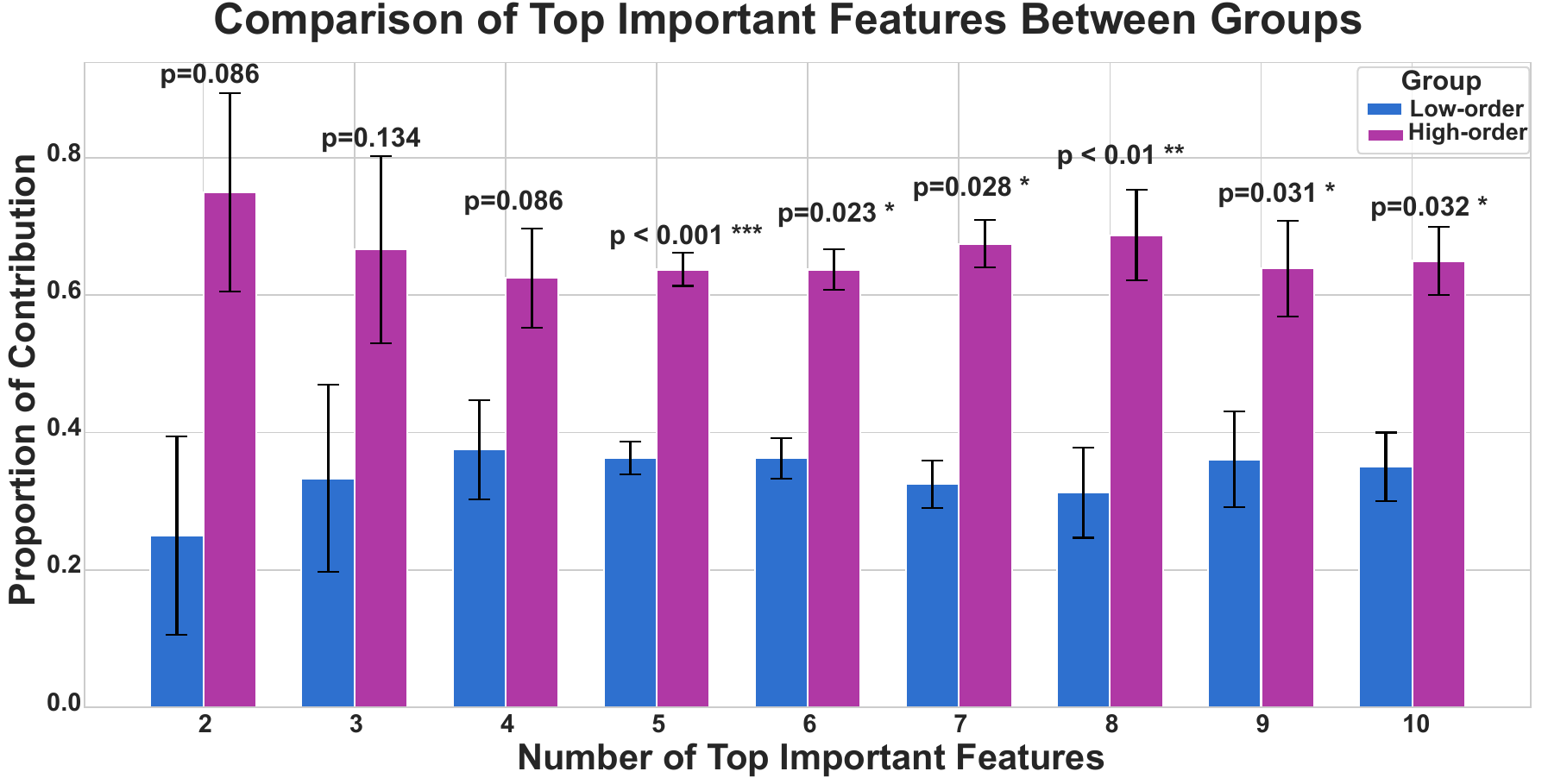}
    \caption{%
    \textbf{Comparison of the contribution of top energy landscape features in predicting working memory performance.}
    Bar plots showing the mean proportion of contribution in the top-$S$ most important features as identified by random forest models. Error bars indicate standard error of the mean across multiple model runs. The $p$-values above each pair represent statistical comparisons between the two groups of feature sets. The high-order group shows a significantly greater contribution to the top ranked features, reflecting stronger discriminative power in relation to working memory performance.}
    \label{fig:feature_importance}
\end{figure}

Lastly, we compared the relative importance of low-order versus high-order energy landscape features using random forest models that were trained on the combined feature sets. The results are provided in Figure~\ref{fig:feature_importance}, which shows the proportion of high versus low-order features among the top-ranked predictors. The high-order energy features were consistently accounted for a greater share of the important predictors. This trend was consistent in top-ranked features across all models, with significant group differences emerging for the top five and higher number of features ($S \geq 5$). In contrast, the low-order energy feature contributions remain comparatively low across all thresholds. These findings support the notion that high-order features of the energy landscape provide more predictive and informative insights into the neural dynamics underlying working memory performance in brain tumor patients.

\section{Discussion}
This study is an important step towards identifying presurgical neural dynamic disruptions that correlate with postoperative cognitive outcome in brain tumor patients. To this end, we introduced high-order energy landscapes, combining subject-specific functional clustering with Maximum Entropy Models (MEM) to characterize presurgical neural dynamics in brain tumor patients and study its relation to postoperative working memory (WM) performance. We have shown that these presurgical high-order neural dynamics are predictive of postoperative working memory outcome.  To the best of our knowledge, this is the first study to relate high-order energy landscape features derived from presurgical rs-fMRI to postoperative working memory outcome in a brain tumor cohort.

\subsection{Interpretation of Energy Landscapes}
We observed a strong association between extreme energy states and postoperative working memory performance in brain tumor patients. More specifically, patients with poor working memory performance demonstrated larger but less frequent energy transitions, whereas patients with high working memory performance showed an opposite trend. Taken together, this suggests the brain operates within a more constrained and energetically demanding functional landscape in brain tumor patients with low working memory performance. The larger energy transitions imply that shifting between cognitive states requires greater neural effort, potentially due to disrupted structural connectivity \cite{roy2025brain}. Meanwhile, the reduced frequency of transitions reflects lower capacity for functional reconfiguration, which is critical for working memory performance \cite{yue2024differential}. Given that energy values reflect stability of brain states, our results suggest that reduced neural stability in brain tumor patients is closely linked to poorer working memory performance, which aligns with recent findings indicating that individuals with more stable task-based functional connectomes achieve better working memory outcomes \cite{corriveau2023functional}. 

\subsection{Applications in Patient Treatment and Management}
If validated in a larger cohort, the findings of this study could facilitate the development of personalized cognitive rehabilitation plans, enabling early intervention to reduce the risk of postoperative cognitive decline. For instance, if a patient is identified as high risk, a proactive neuromodulation strategy could be implemented to enhance neural plasticity and promote functional network reorganization, ultimately improving cognitive outcomes during recovery \cite{ekert2024interventional}. This is particularly important given the unique opportunity to leverage the surgical window to directly modulate neural circuitry that would otherwise be inaccessible \cite{poologaindran2022interventional}. For example, intraoperative neuromodulation techniques such as deep brain stimulation could be employed during tumor resection to support cognitive function and optimize postoperative recovery \cite{boerger2023large}.

More broadly, this study builds upon a growing body of work demonstrating the utility of energy‑landscape based models for characterizing altered brain dynamics across neuropsychiatric and clinical populations. Prior studies have revealed systematic alterations in brain‑state stability, transition structure, and energetic cost in schizophrenia \cite{allen2025energy}, Alzheimer’s disease \cite{xing2024energy}, and substance use disoerdeds \cite{varanasi2026resting}. The present work extends these advances to the presurgical brain tumor domain by introducing a high‑order energy landscape framework that captures system‑level coordination among distributed functional networks. In contrast to prior applications largely focused on diffuse neurodegenerative pathology, brain tumors provide a unique model of focal structural disruption accompanied by large‑scale functional reorganization. Our findings therefore demonstrate how high‑order energy modeling can generalize across clinical contexts and provide mechanistic insight into postoperative cognitive vulnerability by quantifying alterations in the energetic cost, stability, and flexibility of large‑scale brain states.

\subsection{Limitations and Future Directions}
Despite the encouraging results, there are several challenges that need to be addressed to better understand the role of high-order energy landscapes in postoperative cognitive function. From a neurophysiological perspective, although these energy values appear to capture complex high-order neural dynamics, the underlying mechanisms of neural circuitry that give rise to such dynamics remain poorly understood. Incorporating computational models that relate changes in neural circuitry to high-order energy values could provide valuable insights into the biological basis of these patterns \cite{breakspear2017dynamic}. From a methodological perspective, the binarization process during brain state identification causes significant information loss by ignoring subtle changes in brain activity \cite{masuda2025energy}. These subtle variations may be crucial for understanding the continuum nature of brain function and could ultimately hinder the precise characterization of neural dynamics \cite{deco2019awakening}. Hence, exploring alternative approaches based on continuous state-space representations could offer new insights into the underlying high-order neural dynamics and its relation to cognitive performance. Finally, from a clinical perspective, several factors such as inter-subject variability in brain structure and function must be considered before high-order energy metrics can be translated into practical use \cite{seitzman2019trait}. This is particularly important given the pharmacological heterogeneity of patient populations and the various ways in which individual brains may respond to agents such as anesthetics during surgery \cite{luppi2025general}. Hence, personalized approaches and stratified analyses may be necessary to ensure the clinical relevance, robustness, and reliability of these energy-based measures in predicting or monitoring postoperative cognitive outcomes.

\section{Conclusion}

In conclusion, we investigated the relationship between presurgical high-order neural dynamics and postoperative cognitive function in brain tumor patients using energy landscape analysis. Our study focused on the role of extreme energy states, particularly their frequency and magnitude, and their association with postoperative working memory performance. The results revealed that patients with lower cognitive scores exhibit fewer but more extreme energy transitions, suggesting reduced neural flexibility. These findings point to the potential of using high-order energy features as predictive biomarkers for postoperative cognitive outcomes. Nevertheless, more in-depth longitudinal studies are needed to fully understand the role of high-order neuronal dynamics in cognitive function and their long-term implications for individualized treatment strategies and neurorehabilitation.

\section{Conflict of Interest}
None of the authors have potential conflicts of interest to be disclosed.

\section{Ethics Statement}
This study analyzed publicly available datasets and did not involve direct interaction with human participants or animals. Ethical approval and informed consent were not required.

\section{Code and Data Availability}
All the code used in this study are publicly available at:\\
\href{https://github.com/3sigmalab/HODELA}{https://github.com/3sigmalab/HODELA}.
The dataset used in this study is available through \cite{aerts2022pre}.

\section{Acknowledgments}
This research did not receive any specific grant from funding agencies in the public, commercial, or not-for-profit sectors.

\bibliographystyle{elsarticle-num}
\bibliography{refs_abbrev}

\end{document}